\documentstyle[aps,12pt]{revtex}

\begin{document}

\preprint{EFUAZ FT-98-58}

\title{Again on the possible compositeness of the photon.\thanks{Submitted
to ``International Journal Theoretical Physics"}}

\author{{\bf Valeri V. Dvoeglazov}}

\address{Escuela de F\'{\i}sica, Universidad Aut\'onoma de Zacatecas\\
Apartado Postal C-580, Zacatecas 98068 Zac., M\'exico\\
E-mail: valeri@ahobon.reduaz.mx\\
URL: http://ahobon.reduaz.mx/\~~valeri/valeri.htm
}

\date{February, 1998}

\maketitle

\medskip

\begin{abstract}
We construct the transversal and longitudinal 4-vector potentials
and strengths $F^{\mu\nu}$ in the momentum representation on using the
second-type 4-spinors recently proposed by Ahluwalia. Questions of
relevance of this kind of theories to  the correct photon statistics
are briefly discussed.
\end{abstract}

\bigskip

\section{Introduction}

In the papers~\cite{deBr,Jordan,Kronig,Price,Berez,Perkins}
the hypothesis was discussed that a photon is a composite state of
a pair of neutral particles of the $(1/2,0)\oplus (0,1/2)$
representation. Unfortunately, the matters related to the
correct photon statistics have {\it not} been clarified in
the papers on ``the neutrino theory of light".
Recently, after the study of previous papers~\cite{MLC,DVA0}
I asked several questions~\cite{D1} which might be relevant
to this kind of field models. In the present note I give
a formal basis to some of my previous thoughts. Main objections were:
1) why did one use the charged particle (electron-positron)
4-spinors and the Dirac current only in order to construct photon states?
2) why were physicists so unpleasant to accept that the
electromagentic 4-vector  may be longitudinal? 3) why
did the physicists almost not explore the possibility of
using the electric and magnetic fields ${\bf E}$ and ${\bf B}$
in order to describe the light in the quantum field theory?

\section{Classical Fields for a Composite Photon-Notoph}

Here, we first construct the composite photons on
using the 2-nd kind spinors~\cite{DVA,HJ}.
On the basis of the formulas given in the Appendices A and B
one obtains
\begin{mathletters}
\begin{eqnarray}
\overline \lambda^S_\uparrow (p^\mu) \gamma^\mu \lambda^S_\uparrow (p^\mu)
&=& \overline \lambda^A_\uparrow (p^\mu) \gamma^\mu \lambda^A_\uparrow
(p^\mu) =
{m\over N} \left [ u^\mu ({\bf p}, 0_t) - u^\mu ({\bf p}, 0) \right
]\, ,\\
\overline \lambda^S_\downarrow (p^\mu) \gamma^\mu \lambda^S_\downarrow
(p^\mu) &=& \overline \lambda^A_\downarrow (p^\mu) \gamma^\mu
\lambda^A_\downarrow (p^\mu) =
{m\over N} \left [ u^\mu ({\bf p}, 0_t) + u^\mu ({\bf p}, 0)
\right ]\, ,\\
\overline \lambda^S_\uparrow (p^\mu) \gamma^\mu \lambda^S_\downarrow
(p^\mu) &=&
\overline \lambda^S_\downarrow (p^\mu) \gamma^\mu \lambda^S_\uparrow
(p^\mu) =
{m\over N\sqrt{2}} \left [ u^\mu ({\bf p}, +1) - u^\mu ({\bf
p}, -1) \right ]\, ,\\
\overline \lambda^S_\uparrow (p^\mu) \gamma^\mu \lambda^A_\downarrow
(p^\mu) &=& -
\overline \lambda^S_\downarrow (p^\mu) \gamma^\mu \lambda^A_\uparrow
(p^\mu) =
-{m\over N\sqrt{2}} \left [ u^\mu ({\bf p}, +1) + u^\mu ({\bf
p}, -1) \right ]\, .
\end{eqnarray}
\end{mathletters}
After simple alebraic transformations one can write the $(1/2,1/2)$
momentum-space function, for instance, in the following
forms ($a_1+a_2 =1$, $a_3+a_4 = 1$):\footnote{For the sake of the general
consideration it is assumed  the neutrino states to be massive.}
\begin{mathletters}
\begin{eqnarray}
u^\mu ({\bf p}, +1) &=&  {N\over m\sqrt{2}} \left \{
a_1 \left [\overline \lambda^S_\uparrow (p^\mu) \gamma^\mu (1+\gamma^5)
\lambda^S_\downarrow (p^\mu) \right ]  + a_2
\left [\overline \lambda^S_\downarrow (p^\mu) \gamma^\mu (1-\gamma^5)
\lambda^S_\uparrow (p^\mu) \right ]\right \}=\,\nonumber \\
&=&   {N\over m\sqrt{2}} \left \{ a_1 \left [
\overline \lambda^A_\uparrow (p^\mu) \gamma^\mu (1+\gamma^5)
\lambda^A_\downarrow (p^\mu) \right ] +a_2
\left [\overline \lambda^A_\downarrow (p^\mu) \gamma^\mu (1-\gamma^5)
\lambda^A_\uparrow (p^\mu) \right ]\right \}\, , \\
u^\mu ({\bf p}, -1) &=& - {N\over m\sqrt{2}}\left\{ a_3 \left [
\overline \lambda^S_\uparrow (p^\mu) \gamma^\mu (1-\gamma^5)
\lambda^S_\downarrow (p^\mu) \right ] +a_4
\left [\overline \lambda^S_\downarrow (p^\mu) \gamma^\mu (1+\gamma^5)
\lambda^S_\uparrow (p^\mu) \right ] \right \}=\, \nonumber\\
&=&- {N\over m\sqrt{2}} \left \{ a_3\left [
\overline \lambda^A_\uparrow (p^\mu) \gamma^\mu (1-\gamma^5)
\lambda^A_\downarrow (p^\mu) \right ] +a_4
\left [\overline \lambda^A_\downarrow (p^\mu) \gamma^\mu (1+\gamma^5)
\lambda^A_\uparrow (p^\mu) \right ] \right \}\, ,\\
u^\mu ({\bf p}, 0) &=& {N\over 2m} \left [
\overline \lambda^S_\downarrow (p^\mu) \gamma^\mu
\lambda^S_\downarrow (p^\mu) - \overline \lambda^S_\uparrow
(p^\mu) \gamma^\mu \lambda^S_\uparrow (p^\mu) \right ] =\,\nonumber\\
&&\qquad\qquad\qquad =
{N\over 2m} \left [ \overline \lambda^A _\downarrow (p^\mu) \gamma^\mu
\lambda^A_\downarrow (p^\mu) - \overline \lambda^A_\uparrow
(p^\mu) \gamma^\mu \lambda^A_\uparrow (p^\mu) \right ]\,,\\
u^\mu ({\bf p}, 0_t) &=& {N\over 2m} \left [\overline \lambda^S_\uparrow
(p^\mu) \gamma^\mu \lambda^S_\uparrow (p^\mu) +
\overline \lambda^S_\downarrow (p^\mu) \gamma^\mu
\lambda^S_\downarrow (p^\mu)  \right ] =\, \nonumber\\
&&\qquad\qquad\qquad = {N\over 2m}
\left [\overline \lambda^A_\uparrow
(p^\mu) \gamma^\mu \lambda^A_\uparrow (p^\mu) +
\overline \lambda^A _\downarrow (p^\mu) \gamma^\mu
\lambda^A_\downarrow (p^\mu)  \right ]\,.
\end{eqnarray}
\end{mathletters}
Tensor currents may also be expressed by means of the electric/magnetic
strengths:
\begin{mathletters} \begin{eqnarray}
\overline\lambda^S_\uparrow
(p^\mu) \sigma^{\mu\nu} \lambda^S_\downarrow (p^\mu) &=& \overline
\lambda^S_\downarrow (p^\mu) \sigma^{\mu\nu} \lambda^S_\uparrow (p^\mu) =
-{2im \over N} \pmatrix{{\bf E}^{(+)} ({\bf p}, 0)\cr {\bf B}^{(+)} ({\bf
p}, 0)\cr} \, ,\\
\overline\lambda^S_\uparrow (p^\mu) \sigma^{\mu\nu}
\lambda^A_\downarrow (p^\mu) &=& \overline \lambda^S_\downarrow (p^\mu)
\sigma^{\mu\nu} \lambda^A_\uparrow (p^\mu) = {2m \over N} \pmatrix{-{\bf
B}^{(+)} ({\bf p}, 0)\cr {\bf E}^{(+)} ({\bf p}, 0)\cr} = \nonumber\\
&&\qquad\qquad\qquad =i\Gamma^0\Gamma^5 \overline\lambda^S_{\uparrow}
(p^\mu) \sigma^{\mu\nu} \lambda^S_{\downarrow} (p^\mu)
= i\Gamma^0\Gamma^5 \overline\lambda^S_{\downarrow}
(p^\mu) \sigma^{\mu\nu} \lambda^S_{\uparrow} (p^\mu) \, ,\\
&&\nonumber\\
\overline\lambda^S_\uparrow (p^\mu) \sigma^{\mu\nu} \lambda^S_\uparrow
(p^\mu) &=& - {\sqrt{2} im \over N} \pmatrix{
\left [{\bf E}^{(+)} ({\bf p}, +1) - {\bf E}^{(+)} ({\bf p}, -1) \right ]
- i \left [ {\bf B}^{(+)} ({\bf p}, +1) +{\bf B}^{(+)} ({\bf p}, -1)
\right ]\cr
i\left [{\bf E}^{(+)} ({\bf p}, +1) + {\bf E}^{(+)} ({\bf p}, -1) \right ]
+ \left [ {\bf B}^{(+)} ({\bf p}, +1) - {\bf B}^{(+)} ({\bf p}, -1)
\right ]\cr}\nonumber\\
&&\\
&&\nonumber\\
\overline\lambda^S_\uparrow (p^\mu) \sigma^{\mu\nu} \lambda^A_\uparrow
(p^\mu) &=& i\Gamma^0 \Gamma^5 \overline \lambda^S_\uparrow (p^\mu)
\sigma^{\mu\nu} \lambda^S_\uparrow (p^\mu)\, ,\\
&&\nonumber\\
\overline\lambda^S_\downarrow (p^\mu) \sigma^{\mu\nu} \lambda^S_\downarrow
(p^\mu) &=&  {\sqrt{2} im \over N} \pmatrix{
\left [{\bf E}^{(+)} ({\bf p}, +1) - {\bf E}^{(+)} ({\bf p}, -1) \right ]
+ i \left [ {\bf B}^{(+)} ({\bf p}, +1) +{\bf B}^{(+)} ({\bf p}, -1)
\right ]\cr
-i\left [{\bf E}^{(+)} ({\bf p}, +1) + {\bf E}^{(+)} ({\bf p}, -1) \right ]
+ \left [ {\bf B}^{(+)} ({\bf p}, +1) - {\bf B}^{(+)} ({\bf p}, -1)
\right ]\cr}\nonumber\\
&&\\
\overline\lambda^S_\downarrow (p^\mu) \sigma^{\mu\nu} \lambda^A_\downarrow
(p^\mu) &=& i\Gamma^0 \Gamma^5 \overline \lambda^S_\downarrow (p^\mu)
\sigma^{\mu\nu} \lambda^S_\downarrow (p^\mu)\, .
\end{eqnarray}
\end{mathletters}
In the above formulas and below one has $$\Gamma^5 =
\pmatrix{\openone_{3\times 3}& 0\cr 0&-\openone_{3\times 3}\cr}\,,\quad
\Gamma^0 = \pmatrix{0&\openone_{3\times 3}\cr \openone_{3\times 3}&
0\cr}$$.
Hence, we have
\begin{mathletters}
\begin{eqnarray}
{\bf E}^{i\,(+)} ({\bf p}, 0) &=& {iN\over 2m}
\left [ c_1 \overline\lambda^S_\uparrow (p^\mu) \sigma^{0i}
\lambda^S_\downarrow (p^\mu) +c_2
\overline\lambda^S_\downarrow (p^\mu) \sigma^{0i}
\lambda^S_\uparrow (p^\mu) \right ] +\nonumber\\
&+& {N\over 4m} \epsilon^{ijk}
\left [ c_3 \overline\lambda^S_\uparrow (p^\mu) \sigma^{jk}
\lambda^A_\downarrow (p^\mu) + c_4
\overline\lambda^S_\downarrow (p^\mu) \sigma^{jk}
\lambda^A_\uparrow (p^\mu) \right ] \, ,\\
{\bf E}^{(+)} ({\bf p}, +1) &=& c_5 \left \{ {iN\over
4\sqrt{2}m}  \left [ \overline\lambda^S_\uparrow
(p^\mu) \sigma^{0i} \lambda^S_\uparrow (p^\mu) -
\overline\lambda^S_\downarrow (p^\mu) \sigma^{0i}
\lambda^S_\downarrow (p^\mu) \right ]+\right.\nonumber\\
&+& \left.{N\over 8\sqrt{2}m} \epsilon^{ijk}
\left [ \overline\lambda^S_\uparrow (p^\mu) \sigma^{jk}
\lambda^S_\uparrow (p^\mu) +
\overline\lambda^S_\downarrow (p^\mu) \sigma^{jk}
\lambda^S_\downarrow (p^\mu) \right ] \right \}+\nonumber\\
&+&c_6 \left \{ {iN\over
4\sqrt{2}m}  \left [ \overline\lambda^S_\uparrow
(p^\mu) \sigma^{0i} \lambda^A_\uparrow (p^\mu) +
\overline\lambda^S_\downarrow (p^\mu) \sigma^{0i}
\lambda^A_\downarrow (p^\mu) \right ]+\right.\nonumber\\
&+& \left.{N\over 8\sqrt{2}m} \epsilon^{ijk}
\left [ \overline\lambda^S_\uparrow (p^\mu) \sigma^{jk}
\lambda^A_\uparrow (p^\mu) -
\overline\lambda^S_\downarrow (p^\mu) \sigma^{jk}
\lambda^A_\downarrow (p^\mu) \right ] \right \}\, ,\\
{\bf E}^{(+)} ({\bf p}, -1) &=& c_5 \left \{ {-iN\over
4\sqrt{2}m} \left [ \overline\lambda^S_\uparrow
(p^\mu) \sigma^{0i} \lambda^S_\uparrow (p^\mu) -
\overline\lambda^S_\downarrow (p^\mu) \sigma^{0i}
\lambda^S_\downarrow (p^\mu) \right ] + \right .\nonumber\\
&+& \left . {N\over 8\sqrt{2}m} \epsilon^{ijk}
\left [ \overline\lambda^S_\uparrow (p^\mu) \sigma^{jk}
\lambda^S_\uparrow (p^\mu) +
\overline\lambda^S_\downarrow (p^\mu) \sigma^{jk}
\lambda^S_\downarrow (p^\mu) \right ] \right \} +\nonumber\\
&+&c_6 \left \{ {iN\over
4\sqrt{2}m} \left [ \overline\lambda^S_\uparrow
(p^\mu) \sigma^{0i} \lambda^A_\uparrow (p^\mu) +
\overline\lambda^S_\downarrow (p^\mu) \sigma^{0i}
\lambda^A_\downarrow (p^\mu) \right ] - \right .\nonumber\\
&-& \left . {N\over 8\sqrt{2}m} \epsilon^{ijk}
\left [ \overline\lambda^S_\uparrow (p^\mu) \sigma^{jk}
\lambda^A_\uparrow (p^\mu) -
\overline\lambda^S_\downarrow (p^\mu) \sigma^{jk}
\lambda^A_\downarrow (p^\mu) \right ] \right \}\,,\\
{\bf B}^{(+)} ({\bf p}, 0) &=& {iN\over 2m} \epsilon^{ijk}
\left [ c_1 \overline\lambda^S_\uparrow (p^\mu) \sigma^{jk}
\lambda^S_\downarrow (p^\mu) +c_2
\overline\lambda^S_\downarrow (p^\mu) \sigma^{jk}
\lambda^S_\uparrow (p^\mu) \right ] -\nonumber\\
&-& {N\over 4m}
\left [ c_3 \overline\lambda^S_\uparrow (p^\mu) \sigma^{0i}
\lambda^A_\downarrow (p^\mu) +c_4
\overline\lambda^S_\downarrow (p^\mu) \sigma^{0i}
\lambda^A_\uparrow (p^\mu) \right ] \, ,\\
{\bf B}^{(+)} ({\bf p}, +1) &=& c_5 \left \{{-N\over
4\sqrt{2}m} \left [
\overline\lambda^S_\uparrow (p^\mu) \sigma^{0i} \lambda^S_\uparrow
(p^\mu) + \overline\lambda^S_\downarrow (p^\mu) \sigma^{0i}
\lambda^S_\downarrow (p^\mu) \right ] +\right .\nonumber\\
&+& \left . {iN\over 8\sqrt{2}m} \epsilon^{ijk}
\left [ \overline\lambda^S_\uparrow (p^\mu) \sigma^{jk}
\lambda^S_\uparrow (p^\mu) -
\overline\lambda^S_\downarrow (p^\mu) \sigma^{jk}
\lambda^S_\downarrow (p^\mu) \right ] \right \} +\nonumber\\
&+& c_6 \left \{ {-N\over
4\sqrt{2}m} \left [
\overline\lambda^S_\uparrow (p^\mu) \sigma^{0i} \lambda^A_\uparrow
(p^\mu) - \overline\lambda^S_\downarrow (p^\mu) \sigma^{0i}
\lambda^A_\downarrow (p^\mu) \right ] +\right .\nonumber\\
&+& \left . {iN\over 8\sqrt{2}m} \epsilon^{ijk}
\left [ \overline\lambda^S_\uparrow (p^\mu) \sigma^{jk}
\lambda^A_\uparrow (p^\mu) +
\overline\lambda^S_\downarrow (p^\mu) \sigma^{jk}
\lambda^A_\downarrow (p^\mu) \right ] \right \} \, ,\\
{\bf B}^{(+)} ({\bf p}, -1) &=& c_5  \left \{{-N\over
4\sqrt{2}m} \left [
\overline\lambda^S_\uparrow (p^\mu) \sigma^{0i} \lambda^S_\uparrow
(p^\mu) + \overline\lambda^S_\downarrow (p^\mu) \sigma^{0i}
\lambda^S_\downarrow (p^\mu) \right ] - \right .\nonumber\\
&-&\left . {iN\over 8\sqrt{2}m} \epsilon^{ijk}
\left [ \overline\lambda^S_\uparrow (p^\mu) \sigma^{jk}
\lambda^S_\uparrow (p^\mu) -
\overline\lambda^S_\downarrow (p^\mu) \sigma^{jk}
\lambda^S_\downarrow (p^\mu) \right ] \right \} +\nonumber\\
&+&c_6  \left \{ {N\over
4\sqrt{2}m} \left [
\overline\lambda^S_\uparrow (p^\mu) \sigma^{0i} \lambda^A_\uparrow
(p^\mu) - \overline\lambda^S_\downarrow (p^\mu) \sigma^{0i}
\lambda^A_\downarrow (p^\mu) \right ] + \right .\nonumber\\
&+&\left . {iN\over 8\sqrt{2}m} \epsilon^{ijk}
\left [ \overline\lambda^S_\uparrow (p^\mu) \sigma^{jk}
\lambda^A_\uparrow (p^\mu) +
\overline\lambda^S_\downarrow (p^\mu) \sigma^{jk}
\lambda^A_\downarrow (p^\mu) \right ] \right \} \,.
\end{eqnarray} \end{mathletters}
We still apparently note that on using the formulas of the Appendix A
(\ref{t1}-\ref{t8}) one can obtain formally different formulas
connecting an antisymmetric tensor of the rank 2 and tensor currents
composed of $\lambda^{A,S}_{\uparrow\downarrow}$ spinors.
One can also repeat the derivation of the formulas of this paper
on using the Dirac 4-spinors, see the formulas (\ref{c1}-\ref{c4}), thus
arriving at the {\it electron-positron theory of light}. We leave this
simple exercise to the reader.

\section{Where may the Berezinsky arguments fail?}

In refs.~\cite{DVA,HJ,DVO,DVONP} the McLennan-Case construct~\cite{MLC}
has been developed considerably. We showed the theoretical possibility
of existence of {\it bi-orthonormal} states which have slightly
different anticommutation relations comparing with the Dirac fermion.
It was postulated
\begin{mathletters}
\begin{eqnarray}
\left [ a_{ \eta^\prime } ( p^{\prime \, \mu} ), a_\eta^\dagger (p^\mu)
\right ]_\pm
&=& (2\pi)^3 2p_0 \delta^3 ({\bf p} -{\bf p}^\prime )
\delta_{\eta , - \eta^\prime} \, ,\label{cr1}\\
\left [ b_{ \eta^\prime } ( p^{\prime \, \mu} ), b_\eta^\dagger (p^\mu)
\right ]_\pm
&=& (2\pi)^3 2p_0 \delta^3 ({\bf p} -{\bf p}^\prime )
\delta_{\eta , - \eta^\prime } \, , \label{cr2}\\
\left [ a_{\eta^\prime} (p^{\prime\,\mu} ), b_\eta^\dagger (p^\mu)
\right ]_\pm
&=& 0\, \label{cr3}.
\end{eqnarray} \end{mathletters}
for self/anti-self charge
conjugate states.
With taking into account  these new ideas we re-examine the Berezynsky
proof of the Price's theorem.

1) In ref.~\cite{Berez} (particularly in the first postulate, Eq. (6))
the author assumed neutrinos to be massless. In the massive case
we have to use the generalized expressions (see, for instance,
ref.~\cite[p. 177]{Wein}) instead of
(6) of ref.~\cite{Berez}.

2) The author of ref.~\cite{Berez} assumed that neutrinos are fermions
and (in his notation) the operators
$a_i ({\bf k})$ and $b_i ({\bf k})$ obey the commutation relations
$$[ a_i^\dagger ({\bf k}), a_j ({\bf k}^\prime) ]_+
=[ b_i^\dagger ({\bf k}), b_j ({\bf k}^\prime) ]_+ = \delta_{ij}
\delta ( {\bf k} - {\bf k}^\prime ) .$$
If we assume that the neutrinos are {\it bi-orthonormal} states
in the Berezinsky notation the analogues of anticommutation relations
(\ref{cr1}-\ref{cr3}) may be expressed (in a particular case) as follows:
\begin{equation}
\left [ a_i^\dagger ({\bf k}), b_j ({\bf k^\prime}) \right ]_+
= \left [ b_i^\dagger ({\bf k}), a_j ({\bf k^\prime}) \right ]_+
= e^{i\phi (i,j)}\delta_{ij} \delta ({\bf k} - {\bf k^\prime})
\end{equation}
with indices $i$ and $j$ referring to ``$i$-th" or ``$j$-th"
(anti)neutrino, respectively. All other anticommutators are equal
to zero.

3) The right- and left- circular polarized photons have been assumed there
to be massless too. We still advocate that they also may be
self/anti-self charge conjugate (or $\Gamma_5 {\cal C}$
self/anti-self charge conjugate) massive states~\cite{HJ}.

4) As a consequence of the item (3) we suggested~\cite{DEC}
{\it two} definitions of the linear-polarized radiation.

For field theorists it is known that the change of the
polarization state of massive particles can be made by the boost
(and/or other non-unitary operations).
On the other hand, it appears that for $j=1$ states (relevant to the
problem at hand) the change of polarization can be made by means of the
change of the basis of the corresponding {\it complex} vector space, {\it
i.~e.} by the rotation (on the classical level).  It is produced by an
unitary matrix.  For instance, if one describes the magnetic field as
\begin{equation}
{\bf B}^{\mbox{circ.}} =
{B^{(0)}\over \sqrt{2}} \left [\pmatrix{i\cr 1\cr 0\cr} e^{+i\phi} +
\pmatrix{-i\cr 1\cr 0\cr} e^{-i\phi} \right ]\, ,
\end{equation}
($\phi=\omega t -{\bf k}\cdot {\bf r}$)
on using
the unitary matrix
\begin{equation}
U ={1\over \sqrt{2}}\pmatrix{-i&1&0\cr
i&1&0\cr
0& 0&\sqrt{2}}
\end{equation}
one can obtain the linear polarized (in the
plane $XY$) radiation\footnote{If one wishes to see the real-valued
magnetic fields instead of phasors here they are:  \begin{equation} {\bf
B}_x^{\mbox{circ.}} = - \sqrt{2} B^{(0)} \sin\phi \quad , \, {\bf
B}_y^{\mbox{circ.}} = + \sqrt{2} B^{(0)} \cos\phi \quad  \, .
\end{equation}
or
\begin{equation}
{\bf B}_x^{\mbox{lin.}} =
+ B^{(0)} \cos\phi\quad,\, {\bf B}_y^{\mbox{lin.}} = + B^{(0)}
\cos\phi\quad,\,\label{rv}
\end{equation}
{\it i.~e.}, in the latter case one obtains the linear
polarized radiation with the polarization angle
equal to $\pi /4$. Of
course, the given unitary matrix can be easily generalized
to account for other polarization angles.}$^,$\footnote{The
transformation of transverse components
with the matrix $L$ used by G. Hunter~\cite[Eq.(19)]{Hunter}
is {\it not} generally unitary:
\begin{equation} L\sim
\pmatrix{(A-B)\cos\alpha & -(A+B)\sin\alpha & 0\cr
(A-B)\sin\alpha & (A+B)\cos \alpha & 0\cr
0& 0& 1}\, ,
\end{equation}
with $\alpha$  being the polar angle of the cylindrical system of
coordinates. In the case of the linear polarization defined
in such a way~\cite{Hunter} one has ${\bf B}\times {\bf B}^\ast =
0$.  This transformation may also change the normalization of the
corresponding vectors which in the quantized case correspond to a particle
and an anti-particle. The determinant
of the transformation is, in general, {\it not} equal to the unit.  While
the determinant of our matrix is also not equal to the unit
($\mbox{det} U = -i$), but the norm of the corresponding
quantum states is still preserved (while this is not so for the
corresponding real quantities).}
\begin{equation} {\bf B}^{\mbox{lin.}} = U {\bf B}^{\mbox{circ.}} =
B^{(0)} \left [ \pmatrix{1\cr 0\cr 0\cr} e^{+i\phi}+\pmatrix{0\cr 1\cr
0\cr} e^{-i\phi} \right ]\ .
\end{equation}
Of course, the corresponding quantum covering of the above transformation
will be different from (9) of ref.~\cite{Berez}.

5) Furthermore, in private communication in January, 1999
Prof. W. Perkins pointed out~\cite{Perk2} that the requirement of
what Berezinsky  considered as a condition of ``genuinely neutrality" for
a photon (the condition 5 of Berezinsky, Eqs. (10,11) of
ref.~\cite{Berez}) may be lifted (see also [6b]).

\section{Discussion and Conclusion}

In the papers of Barut, {\it e.g.}, ref.~\cite{Barut}
a self-field formulation of quantum
electrodynamics have been proposed. It is based
on the use of the solution
\begin{equation}\label{pot}
{\cal A}^\mu (x) = \int d^4 y D^{\mu\nu} (x-y) j_\nu (y)
\end{equation}
of the coupled Maxwell-Dirac equation
\begin{equation}
\partial^\mu F_{\mu\nu} (x) = e \overline{\Psi} (x) \gamma_\nu \Psi
(x)\quad.
\end{equation}
$D_{\mu\nu} (x-y)$  is a Green's function of
electromagnetic field in the usual potential formulation.
In a series of the works A. Barut {\it et al.} have shown that this
formulation of quantum electrodynamics (based on the iteration
procedure, {\it not} on the perturbation theory) leads to the {\it same}
experimental predictions as the ordinary formalism.

Let me try to write the formula  (\ref{pot}) in the momentum space
(To my knowledge, such attempts are absent in the literature).
I consider  momenta as $q=\lambda t$ and $p= (\lambda -1) t$,
$\lambda$ is some function spanned from $0$  to $1$.
In this case
\begin{eqnarray}
{\cal A}_\mu (x) &=& -e\int {d^3 {\bf p} d^3 {\bf q} \over
(2\pi)^6} \frac{D_F ((q-p)^2)}{2m \sqrt{E_p E_q}}
\sum_{\sigma\sigma^\prime}\left \{ \overline u_\sigma ({\bf p})
\gamma^\mu u_{\sigma^\prime} ({\bf q}) e^{i (q-p) x} a^\dagger_\sigma
({\bf p}) a_{\sigma^\prime} ({\bf q})+ \right .\nonumber\\
&+&\left . \overline v_\sigma ({\bf p}) \gamma^\mu v_{\sigma^\prime}
({\bf q}) e^{-i (q-p) x} b_\sigma ({\bf p})
b^\dagger_{\sigma^\prime} ({\bf q})\right \} \quad,
\end{eqnarray}
and, hence,
\begin{equation}
{\cal A}_\mu (t) = \int_0^1 d\lambda f (\lambda,
t^2 ) \sum_{\sigma\sigma^\prime,\,\pm} \overline\psi^{\pm}_\sigma
((\lambda -1)t) \gamma^\mu \psi^{\pm}_{\sigma^\prime} (\lambda t) \quad.
\end{equation}
Surprisingly, you may see the
well-known Jordan {\it ansatz}.  Thus, referring to the remark of the
previous paragraph one can state the longitudinal de Broglie-Jordan-Barut
potential can describe quantumelectrodynamic processes sufficiently good.

We think that in order to describe  transverse components of the
4-vector potential (left- and right- polarized radiations) correctly one
should set up the different commutation relations for the 4-spinor
fields which are different from those used in the Dirac theory. It is also
possible that in order to overcome dificulties related to teh Pryce
theorem one should use the generalized definition of linear polarized
radiation.  In general, the room of choosing the corresponding constants
in the superpositions permits one to obtain various types of
(anti)commutation relations for the composite particles Tensor currents
can also be generalzied.  In my opinion, resulting expressions of the type
$F^{\mu\nu}_\lambda \sim \bar \nu_\eta (\sigma^{\mu\nu} \pm i \widetilde
\sigma^{\mu\nu}) \nu_{\eta^\prime}$ (and its dual conjugates) also permit
to construct the neutrino theory of light.

{\it Acknowledgments.} The work was motivated by
very useful frank discussions and phone  conversations with Prof. A. F.
Pashkov during last 15 years.  I acknowledge email communications from
Profs. D. V. Ahluwalia, V. Berezinsky, D. Bernard, B. Fauser, W. Perkins
and J. R. Zeni.

Zacatecas University, M\'exico, is thanked for awarding
the full professorship.  This work has been partly supported by
the Mexican Sistema Nacional de Investigadores.

\section*{Appendix A}

Explicit forms of the second-type 4-spinors, refs.~\cite{DVA,HJ,DVO}
in the Weyl representation are (the spinorial basis is fixed
as in ref.~\cite{DVO}, cf.~\cite{DVONP,DVOSP}):
\begin{mathletters}
\begin{eqnarray}
&&\lambda^S_\uparrow (p^\mu) = +i\rho^A_\downarrow
(p^\mu) ={1\over 2 \sqrt{p_0+m}}
\pmatrix{ip_l\cr i(p^- +m)\cr p^- +m\cr -p_r\cr}\, ,\,\,\\
&&\lambda^A_\uparrow (p^\mu) = -i\rho^S_\downarrow (p^\mu) =
{1\over 2 \sqrt{p_0+m}}
\pmatrix{-ip_l\cr -i(p^- +m)\cr p^- +m\cr -p_r\cr}\,,\,\, \\
&&\lambda^S_\downarrow (p^\mu) = -i\rho^A_\uparrow (p^\mu)
= {1\over 2\sqrt{p_0+m}} \pmatrix{-i(p^+ + m)\cr
-ip_r\cr -p_l\cr p^+ +m\cr}\,,\,\,\\
&&\lambda^A_\downarrow (p^\mu) = +i\rho^S_\uparrow (p^\mu) = {1\over 2
\sqrt{p_0+m}} \pmatrix{i(p^+ + m)\cr ip_r\cr -p_l\cr p^+ +m\cr}\quad .
\end{eqnarray}
\end{mathletters}
($p_{r,l} = p_x \pm ip_y$, $p^\pm =p_0 \pm p_z$).

Their connections with the Dirac 4-spinors are the following:
\begin{mathletters}
\begin{eqnarray}
\lambda^S_\uparrow (p^\mu) &=& = +i {1+\gamma^5 \over 2} u_\downarrow
(p^\mu)  +{1 - \gamma^5 \over 2} u_\uparrow (p^\mu)\quad,\label{c1}\\
\lambda^S_\downarrow (p^\mu) &=& = -i {1+\gamma^5 \over 2} u_\uparrow
(p^\mu)  +{1 - \gamma^5 \over 2} u_\downarrow (p^\mu)\quad,\\
\lambda^A_\uparrow (p^\mu) &=& = -i {1+\gamma^5 \over 2} u_\downarrow
(p^\mu)  +{1 - \gamma^5 \over 2} u_\uparrow (p^\mu)\quad,\\
\lambda^A_\downarrow (p^\mu) &=& = +i {1+\gamma^5 \over 2} u_\uparrow
(p^\mu)  +{1 - \gamma^5 \over 2} u_\downarrow (p^\mu)\quad.\label{c4}
\end{eqnarray}
\end{mathletters}

Normalization conditions read:
\begin{mathletters}
\begin{eqnarray}
\overline \lambda^S_\uparrow (p^\mu) \lambda^S_\downarrow (p^\mu)
&=& - im\,,
\quad\overline \lambda^S_\downarrow (p^\mu) \lambda^S_\uparrow (p^\mu)
= + im\quad,\label{n1}\\
\overline \lambda^A_\uparrow (p^\mu) \lambda^A_\downarrow (p^\mu)
&=& + im\,,
\quad\overline \lambda^A_\downarrow (p^\mu) \lambda^A_\uparrow (p^\mu)
= - im\quad.\label{n4}
\end{eqnarray}
\end{mathletters}
All other products for $\lambda$- spinors are equal to zero.

From (\ref{n1},\ref{n4}) one deduces:
\begin{mathletters}
\begin{eqnarray}
\overline \lambda^S_\uparrow (p^\mu) \gamma^5 \lambda^A_\downarrow (p^\mu)
&=& + im\,,\quad
\overline \lambda^S_\downarrow (p^\mu) \gamma^5 \lambda^A_\uparrow (p^\mu)
= - im\,.\\
\overline \lambda^A_\uparrow (p^\mu) \gamma^5 \lambda^S_\downarrow (p^\mu)
&=& - im\,,\quad
\overline \lambda^A_\downarrow (p^\mu) \gamma^5 \lambda^S_\uparrow (p^\mu)
= + im\, .
\end{eqnarray}
\end{mathletters}
All other products for $\lambda$- spinors are equal to zero.

Vector currents are:\footnote{We imply  that the first element
corresponds to $\mu=0$ and, subsequently, $\mu=1,2,3$.}
\begin{mathletters}
\begin{eqnarray}
\overline \lambda^S_\uparrow (p^\mu) \gamma^\mu \lambda^S_\uparrow (p^\mu)
=\pmatrix{p^-\cr p_1 -{p_1 p_3 \over p_0 +m}\cr
p_2 - {p_2 p_3 \over p_0 +m}\cr
p_3 - m -{p_3^2 \over p_0+m}\cr}\,,\quad
\overline \lambda^S_\uparrow (p^\mu) \gamma^\mu \lambda^S_\downarrow
(p^\mu) = - \pmatrix{p_1\cr m + {p_1^2 \over p_0 +m}\cr {p_1 p_2
\over p_0 +m}\cr {p_1 p_3 \over p_0+m}\cr}\,,\\
&&\\
\overline \lambda^S_\uparrow (p^\mu) \gamma^\mu \lambda^A_\uparrow (p^\mu)
=\pmatrix{0\cr 0\cr 0\cr 0\cr}\,,\quad
\overline \lambda^S_\uparrow (p^\mu) \gamma^\mu \lambda^A_\downarrow
(p^\mu) = +i\pmatrix{p_2\cr {p_1 p_2 \over p_0 +m}\cr m+ {p_2^2
\over p_0 +m}\cr {p_2 p_3 \over p_0+m}\cr}\,,\\
&&\\
\overline \lambda^S_\downarrow (p^\mu) \gamma^\mu \lambda^S_\downarrow
(p^\mu) =\pmatrix{p^+\cr p_1 +{p_1 p_3 \over p_0 +m}\cr p_2 + {p_2 p_3
\over p_0 +m}\cr p_3 + m + {p_3^2 \over p_0+m}\cr}\,,\quad
\overline \lambda^S_\downarrow (p^\mu) \gamma^\mu \lambda^S_\uparrow
(p^\mu) = - \pmatrix{p_1\cr m + {p_1^2 \over p_0 +m}\cr {p_1 p_2 \over p_0
+m}\cr {p_1 p_3 \over p_0+m}\cr}\,,\\
&&\\
\overline \lambda^S_\downarrow
(p^\mu) \gamma^\mu \lambda^A_\downarrow (p^\mu) =\pmatrix{0\cr 0\cr 0\cr
0\cr}\,,\quad \overline \lambda^S_\downarrow (p^\mu) \gamma^\mu
\lambda^A_\uparrow (p^\mu) = -i\pmatrix{p_2\cr {p_1 p_2 \over p_0 +m}\cr
m+ {p_2^2 \over p_0 +m}\cr {p_2 p_3 \over p_0+m}\cr}\,;
\end{eqnarray}
\end{mathletters}
and, hence,
\begin{mathletters}
\begin{eqnarray}
\overline \lambda^A_\uparrow (p^\mu) \gamma^\mu \lambda^S_\uparrow (p^\mu)
&=& 0\quad,\\
\overline \lambda^A_\uparrow (p^\mu) \gamma^\mu \lambda^S_\downarrow
(p^\mu) &=& \overline\lambda^S_\uparrow (p^\mu) \gamma^\mu
\lambda^A_\downarrow (p^\mu)\quad,\\
\overline \lambda^A_\uparrow (p^\mu)
\gamma^\mu \lambda^A_\uparrow (p^\mu) &=& \overline\lambda^S_\uparrow
(p^\mu) \gamma^\mu \lambda^S_\uparrow (p^\mu)\quad,\\
\overline \lambda^A_\uparrow (p^\mu) \gamma^\mu
\lambda^A_\downarrow (p^\mu) &=&
\overline\lambda^S_\uparrow (p^\mu) \gamma^\mu \lambda^S_\downarrow
(p^\mu)\quad,\\
\overline
\lambda^A_\downarrow (p^\mu) \gamma^\mu \lambda^S_\downarrow (p^\mu) &=&
0\quad,\\
\overline \lambda^A_\downarrow (p^\mu) \gamma^\mu
\lambda^S_\uparrow (p^\mu) &=& \overline\lambda^S_\downarrow (p^\mu)
\gamma^\mu \lambda^A_\uparrow (p^\mu)\quad,\\
\overline \lambda^A_\downarrow (p^\mu) \gamma^\mu
\lambda^A_\uparrow (p^\mu) &=& \overline\lambda^S_\downarrow (p^\mu)
\gamma^\mu \lambda^S_\uparrow (p^\mu)\quad,\\
\overline
\lambda^A_\downarrow (p^\mu) \gamma^\mu \lambda^A_\downarrow (p^\mu) &=&
\overline\lambda^S_\downarrow (p^\mu) \gamma^\mu \lambda^S_\downarrow
(p^\mu)\quad.
\end{eqnarray} \end{mathletters}

Axial-vector currents are:
\begin{mathletters}
\begin{eqnarray}
\overline \lambda^S_\uparrow (p^\mu) \gamma^\mu \gamma^5
\lambda^S_\uparrow (p^\mu) = \pmatrix{0\cr 0\cr 0\cr 0\cr}\,,\quad
\overline \lambda^S_\uparrow (p^\mu) \gamma^\mu \gamma^5
\lambda^S_\downarrow (p^\mu) =  -i\pmatrix{p_2\cr {p_1
p_2 \over p_0 +m}\cr m+ {p_2^2 \over p_0 +m}\cr {p_2 p_3 \over
p_0+m}\cr}\,,\\
&&\\
\overline \lambda^S_\uparrow (p^\mu) \gamma^\mu \gamma^5
\lambda^A_\uparrow (p^\mu) = - \pmatrix{p^-\cr p_1 -{p_1 p_3 \over p_0
+m}\cr p_2 - {p_2 p_3 \over p_0 +m}\cr p_3 - m -{p_3^2 \over
p_0+m}\cr}\,,\quad
\overline \lambda^S_\uparrow (p^\mu) \gamma^\mu \gamma^5
\lambda^A_\downarrow (p^\mu) =  \pmatrix{p_1\cr m + {p_1^2 \over p_0
+m}\cr {p_1 p_2 \over p_0 +m}\cr {p_1 p_3 \over p_0+m}\cr}\,,\\
&&\\
\overline \lambda^S_\downarrow (p^\mu) \gamma^\mu \gamma^5
\lambda^S_\downarrow (p^\mu) = \pmatrix{0\cr 0\cr 0\cr 0\cr}\,,\quad
\overline \lambda^S_\downarrow (p^\mu) \gamma^\mu \gamma^5
\lambda^S_\uparrow (p^\mu) = + i\pmatrix{p_2\cr {p_1 p_2
\over p_0 +m}\cr m+ {p_2^2 \over p_0 +m}\cr {p_2 p_3 \over
p_0+m}\cr}\,,\\
&&\\
\overline \lambda^S_\downarrow (p^\mu) \gamma^\mu \gamma^5
\lambda^A_\uparrow (p^\mu) = \pmatrix{p_1\cr m + {p_1^2 \over p_0 +m}\cr
{p_1 p_2 \over p_0 +m}\cr {p_1 p_3 \over p_0+m}\cr}\,,\quad
\overline \lambda^S_\downarrow (p^\mu) \gamma^\mu \gamma^5
\lambda^A_\downarrow (p^\mu) = - \pmatrix{p^+\cr p_1 +{p_1 p_3 \over p_0
+m}\cr p_2 + {p_2 p_3 \over p_0 +m}\cr p_3 + m + {p_3^2 \over
p_0+m}\cr}\,;
\end{eqnarray} \end{mathletters}
and, hence,
\begin{mathletters}
\begin{eqnarray}
\overline \lambda^A_\uparrow (p^\mu) \gamma^\mu
\gamma^5 \lambda^S_\uparrow (p^\mu)
&=& \overline\lambda^S_\uparrow (p^\mu) \gamma^\mu \gamma^5
\lambda^A_\uparrow  (p^\mu)\quad,\\
\overline \lambda^A_\uparrow (p^\mu) \gamma^\mu
\gamma^5 \lambda^S_\downarrow
(p^\mu) &=& \overline\lambda^S_\uparrow (p^\mu) \gamma^\mu
\gamma^5 \lambda^A_\downarrow (p^\mu)\quad,\\
\overline \lambda^A_\uparrow (p^\mu) \gamma^\mu \gamma^5
\lambda^A_\uparrow (p^\mu) &=& 0\quad,\\
\overline \lambda^A_\uparrow (p^\mu) \gamma^\mu
\gamma^5 \lambda^A_\downarrow (p^\mu) &=&
\overline\lambda^S_\uparrow (p^\mu)
\gamma^\mu \gamma^5 \lambda^S_\downarrow (p^\mu)\quad,\\
\overline \lambda^A_\downarrow (p^\mu) \gamma^\mu
\gamma^5 \lambda^S_\downarrow (p^\mu) &=&
\overline \lambda^S_\downarrow (p^\mu) \gamma^\mu \gamma^5
\lambda^A_\downarrow (p^\mu) \\
\overline \lambda^A_\downarrow (p^\mu) \gamma^\mu
\gamma^5 \lambda^S_\uparrow (p^\mu) &=&
\overline\lambda^S_\downarrow (p^\mu)
\gamma^\mu \gamma^5 \lambda^A_\uparrow (p^\mu)\quad,\\
\overline \lambda^A_\downarrow (p^\mu) \gamma^\mu
\gamma^5 \lambda^A_\downarrow (p^\mu) &=& 0\quad,\\
\overline \lambda^A_\downarrow (p^\mu) \gamma^\mu
\gamma^5 \lambda^A_\uparrow (p^\mu) &=&
\overline\lambda^S_\downarrow (p^\mu)
\gamma^\mu \gamma^5 \lambda^S_\uparrow (p^\mu)\quad.
\end{eqnarray}
\end{mathletters}

Finally, tensor currents are:\footnote{We imply that the first
element corresponds to $\mu=0,\nu=1$, and, subsequently,
$(\mu=0,\nu=2)$, $(\mu=0, \nu=3)$, $(\mu=2,\nu=3)$, $(\mu=3, \nu=1)$
$(\mu=1,\nu=2)$.}
\begin{mathletters}
\begin{eqnarray}
\overline \lambda^S_\uparrow (p^\mu) \sigma^{\mu\nu}
\lambda^S_\uparrow (p^\mu) &=& \pmatrix{-p^- +{p_1^2 \over p_0 +m}\cr
{p_1 p_2 \over p_0 +m}\cr -p_1 +{p_1 p_3 \over p_0 +m}\cr
-{p_1 p_2 \over p_0 +m}\cr
p^- -{p_2^2 \over p_0 +m}\cr
p_2 -{p_2 p_3 \over p_0 +m}\cr }\quad,\\
\overline \lambda^S_\uparrow (p^\mu) \sigma^{\mu\nu} \lambda^S_\downarrow
(p^\mu) &=& \overline \lambda^S_\downarrow (p^\mu) \sigma^{\mu\nu}
\lambda^S_\uparrow (p^\mu) =
\pmatrix{-{p_1 p_3 \over p_0+m}\cr -{p_2 p_3 \over p_0+m}\cr
p_0 - {p_3^2 \over p_0 +m}\cr p_2\cr -p_1\cr 0\cr}\quad,\\
\overline \lambda^S_\uparrow (p^\mu) \sigma^{\mu\nu} \lambda^A_\uparrow
(p^\mu) &=& +i \pmatrix{{p_1 p_2 \over p_0+m}\cr -p^- + {p_2^2 \over
p_0+m}\cr -p_2 + {p_2 p_3 \over p_0 +m}\cr
- p^- + {p_1^2 \over p_0+m}\cr{p_1 p_2 \over p_0 +m}\cr
-p_1 +{p_1 p_3 \over p_0 +m}\cr}\quad,\\
\overline \lambda^S_\uparrow (p^\mu) \sigma^{\mu\nu}
\lambda^A_\downarrow (p^\mu) &=&  \overline \lambda^S_\downarrow (p^\mu)
\sigma^{\mu\nu} \lambda^A_\uparrow (p^\mu) =
+i \pmatrix{-p_2 \cr p_1 \cr 0\cr
-{p_1 p_3\over
p_0 +m}\cr - {p_2 p_3 \over p_0 +m}\cr
p_0 -{p_3^2 \over p_0 +m}\cr }\quad,\\
\overline \lambda^S_\downarrow (p^\mu) \sigma^{\mu\nu}
\lambda^S_\downarrow (p^\mu) &=& - \pmatrix{-p^+ +{p_1^2 \over p_0 +m}\cr
{p_1 p_2 \over p_0 +m}\cr p_1 +{p_1 p_3 \over p_0 +m}\cr
{p_1 p_2 \over p_0 +m}\cr -p^+ +{p_2^2 \over p_0 +m}\cr
p_2 +{p_2 p_3 \over p_0 +m}\cr }\quad,\\
\overline \lambda^S_\downarrow (p^\mu) \sigma^{\mu\nu}
\lambda^A_\downarrow (p^\mu) &=& +i \pmatrix{{p_1 p_2 \over p_0+m}\cr -p^+
+ {p_2^2 \over p_0+m}\cr p_2 + {p_2 p_3 \over p_0 +m}\cr
p^+ - {p_1^2 \over p_0+m}\cr -{p_1 p_2 \over p_0 +m}\cr
-p_1 -{p_1 p_3 \over p_0 +m}\cr}\quad.
\end{eqnarray}
\end{mathletters}
and, hence,
\begin{mathletters}
\begin{eqnarray}
\overline \lambda^A_\uparrow (p^\mu) \sigma^{\mu\nu}
\lambda^S_\uparrow (p^\mu) &=& -
\overline \lambda^S_\uparrow (p^\mu) \sigma^{\mu\nu}
\lambda^A_\uparrow (p^\mu)\quad,\label{t1}\\
\overline \lambda^A_\uparrow (p^\mu) \sigma^{\mu\nu}
\lambda^S_\downarrow (p^\mu) &=& -
\overline \lambda^S_\uparrow (p^\mu) \sigma^{\mu\nu}
\lambda^A_\downarrow (p^\mu)\quad,\\
\overline \lambda^A_\uparrow (p^\mu) \sigma^{\mu\nu}
\lambda^A_\uparrow (p^\mu) &=& -
\overline \lambda^S_\uparrow (p^\mu) \sigma^{\mu\nu}
\lambda^S_\uparrow (p^\mu)\quad,\\
\overline \lambda^A_\uparrow (p^\mu) \sigma^{\mu\nu}
\lambda^A_\downarrow (p^\mu) &=& -
\overline \lambda^S_\uparrow (p^\mu) \sigma^{\mu\nu}
\lambda^S_\downarrow (p^\mu)\quad,\\
\overline \lambda^A_\downarrow (p^\mu) \sigma^{\mu\nu}
\lambda^S_\uparrow (p^\mu) &=& -
\overline \lambda^S_\downarrow (p^\mu) \sigma^{\mu\nu}
\lambda^A_\uparrow (p^\mu)\quad,\\
\overline \lambda^A_\downarrow (p^\mu) \sigma^{\mu\nu}
\lambda^S_\downarrow (p^\mu) &=& -
\overline \lambda^S_\downarrow (p^\mu) \sigma^{\mu\nu}
\lambda^A_\downarrow (p^\mu)\quad,\\
\overline \lambda^A_\downarrow (p^\mu) \sigma^{\mu\nu}
\lambda^A_\uparrow (p^\mu) &=& -
\overline \lambda^S_\downarrow (p^\mu) \sigma^{\mu\nu}
\lambda^S_\uparrow (p^\mu)\quad,\\
\overline \lambda^A_\downarrow (p^\mu) \sigma^{\mu\nu}
\lambda^A_\downarrow (p^\mu) &=& -
\overline \lambda^S_\downarrow (p^\mu) \sigma^{\mu\nu}
\lambda^S_\downarrow (p^\mu)\quad.\label{t8}
\end{eqnarray}
\end{mathletters}

One can write corresponding relations between currents
composed from the first-type spinors and 4-vector potentials
and electromagnetic strengths. They are found either by direct
calculations or by the application of the formulas of this Appendix.

Normalizations of the Dirac spinors (see the explicit forms
in~\cite[p.53]{Ryder}\footnote{Those 4-spinors were given in the
standard representation. Please pay attention to misprints
in formulas (2.137,2.138) of ref.~\cite{Ryder},
${E+m \over 2m} \Rightarrow \sqrt{{E+m \over 2m}}$.}) are
\begin{mathletters}
\begin{eqnarray}
\overline u_\uparrow (p^\mu) u_\uparrow (p^\mu) = m\, \quad
\overline u_\downarrow (p^\mu) u_\downarrow (p^\mu) = m\, ,\\
\overline v_\uparrow (p^\mu) v_\uparrow (p^\mu) = - m\, \quad
\overline v_\downarrow (p^\mu) v_\downarrow (p^\mu) = - m\, .
\end{eqnarray}
\end{mathletters}
Therefore, one has
\begin{mathletters}
\begin{eqnarray}
\overline u_\uparrow (p^\mu) \gamma^5 v_\uparrow (p^\mu) = m\, \quad
\overline u_\downarrow (p^\mu) \gamma^5 v_\downarrow (p^\mu) = m\, ,
\\ \overline v_\uparrow (p^\mu) \gamma^5 u_\uparrow (p^\mu) = - m\,
\quad \overline v_\downarrow (p^\mu) \gamma^5 u_\downarrow (p^\mu) = - m\,
; \end{eqnarray} \end{mathletters}
and
\begin{mathletters}
\begin{eqnarray}
&&\overline u_\uparrow (p^\mu) \gamma^\mu u_\uparrow (p^\mu) =
\overline u_\downarrow (p^\mu) \gamma^\mu u_\downarrow (p^\mu) =\\
&&\qquad\qquad
=\overline u_\uparrow (p^\mu) \gamma^\mu \gamma^5 v_\uparrow (p^\mu) =
\overline u_\downarrow (p^\mu) \gamma^\mu \gamma^5 v_\downarrow (p^\mu)
= {m\over N} u^\mu ({\bf p}, 0_t)\, ,\\
&&\overline u_\uparrow (p^\mu) \gamma^\mu v_\uparrow (p^\mu) =
-\overline u_\downarrow (p^\mu) \gamma^\mu v_\downarrow (p^\mu) =\\
&&\qquad\qquad\qquad
=\overline u_\uparrow (p^\mu) \gamma^\mu \gamma^5 u_\uparrow (p^\mu) =
-\overline u_\downarrow (p^\mu) \gamma^\mu \gamma^5 u_\downarrow (p^\mu) =
{m\over N} u^\mu ({\bf p}, 0)\, , \\
&&\overline u_\uparrow (p^\mu) \gamma^\mu u_\downarrow (p^\mu)
=\overline u_\downarrow (p^\mu) \gamma^\mu u_\uparrow (p^\mu) =
\overline u_\uparrow (p^\mu) \gamma^\mu \gamma^5 v_\downarrow (p^\mu)
=\overline u_\downarrow (p^\mu) \gamma^\mu \gamma^5 v_\uparrow (p^\mu) =
 0\, ,\\
&&\overline u_\uparrow (p^\mu) \gamma^\mu v_\downarrow (p^\mu) =
\overline u_\uparrow (p^\mu) \gamma^\mu \gamma^5 u_\downarrow (p^\mu) =
{m\sqrt{2}\over N} u^\mu ({\bf p}, -1)\, , \\
&&\overline u_\downarrow (p^\mu) \gamma^\mu v_\uparrow (p^\mu) =
\overline u_\downarrow (p^\mu) \gamma^\mu u_\uparrow (p^\mu)
= - {m\sqrt{2}\over N} u^\mu ({\bf p}, +1)\, . \quad
\end{eqnarray}
\end{mathletters}
One also deduces
\begin{mathletters}
\begin{eqnarray}
&&\overline v_\uparrow (p^\mu) \gamma^\mu u_\uparrow (p^\mu) =
\overline u_\uparrow (p^\mu) \gamma^\mu v_\uparrow (p^\mu) \, ,\\
&&\overline v_\uparrow (p^\mu) \gamma^\mu u_\downarrow (p^\mu) =
\overline u_\uparrow (p^\mu) \gamma^\mu v_\downarrow (p^\mu) \, ,\\
&&\overline v_\uparrow (p^\mu) \gamma^\mu v_\uparrow (p^\mu) =
\overline u_\uparrow (p^\mu) \gamma^\mu u_\uparrow (p^\mu) \, ,\\
&&\overline v_\uparrow (p^\mu) \gamma^\mu v_\downarrow (p^\mu) =
\overline u_\uparrow (p^\mu) \gamma^\mu u_\downarrow (p^\mu) = 0\, ,\\
&&\overline v_\downarrow (p^\mu) \gamma^\mu u_\uparrow (p^\mu) =
\overline u_\downarrow (p^\mu) \gamma^\mu v_\uparrow (p^\mu) \, ,\\
&&\overline v_\downarrow (p^\mu) \gamma^\mu u_\downarrow (p^\mu) =
\overline u_\downarrow (p^\mu) \gamma^\mu v_\downarrow (p^\mu) \, ,\\
&&\overline v_\downarrow (p^\mu) \gamma^\mu v_\uparrow (p^\mu) =
\overline u_\downarrow (p^\mu) \gamma^\mu u_\uparrow (p^\mu) =0 \, ,\\
&&\overline v_\downarrow (p^\mu) \gamma^\mu v_\downarrow (p^\mu) =
\overline u_\downarrow (p^\mu) \gamma^\mu u_\downarrow (p^\mu) \, ;
\end{eqnarray}
\end{mathletters}
and
\begin{mathletters}
\begin{eqnarray}
&&\overline v_\uparrow (p^\mu) \gamma^\mu \gamma^5 u_\uparrow (p^\mu) =
\overline u_\uparrow (p^\mu) \gamma^\mu \gamma^5 v_\uparrow (p^\mu) \, ,\\
&&\overline v_\uparrow (p^\mu) \gamma^\mu \gamma^5 u_\downarrow (p^\mu) =
\overline u_\uparrow (p^\mu) \gamma^\mu \gamma^5 v_\downarrow (p^\mu) =0\,
,\\
&&\overline v_\uparrow (p^\mu) \gamma^\mu \gamma^5 v_\uparrow (p^\mu)
= \overline u_\uparrow (p^\mu) \gamma^\mu \gamma^5 u_\uparrow (p^\mu) \,
,\\
&&\overline v_\uparrow (p^\mu) \gamma^\mu \gamma^5 v_\downarrow
(p^\mu) = \overline u_\uparrow (p^\mu) \gamma^\mu \gamma^5 u_\downarrow
(p^\mu) \, ,\\
&&\overline v_\downarrow (p^\mu) \gamma^\mu \gamma^5
u_\uparrow (p^\mu) = \overline u_\downarrow (p^\mu) \gamma^\mu \gamma^5
v_\uparrow (p^\mu) =0\, , \\
&&\overline v_\downarrow (p^\mu) \gamma^\mu
\gamma^5 u_\downarrow (p^\mu) = \overline u_\downarrow (p^\mu) \gamma^\mu
\gamma^5 v_\downarrow (p^\mu) \, ,\\
&&\overline v_\downarrow (p^\mu)
\gamma^\mu \gamma^5 v_\uparrow (p^\mu) = \overline u_\downarrow (p^\mu)
\gamma^\mu \gamma^5 u_\uparrow (p^\mu) \, ,\\
&&\overline v_\downarrow
(p^\mu) \gamma^\mu \gamma^5 v_\downarrow (p^\mu) = \overline u_\downarrow
(p^\mu) \gamma^\mu \gamma^5 u_\downarrow (p^\mu) \, .
\end{eqnarray}
\end{mathletters}
Finally,
\begin{mathletters} \begin{eqnarray} \overline
u_\uparrow (p^\mu) \sigma^{\mu\nu} u_\uparrow (p^\mu) = {2im \over N}
\pmatrix{ {\bf B}^{(+)} ({\bf p}, 0)\cr - {\bf E}^{(+)} ({\bf p}, 0)\cr}\,
,\label{tci} \\ \overline u_\uparrow (p^\mu) \sigma^{\mu\nu} v_\uparrow
(p^\mu) = {2m \over N} \pmatrix{ {\bf E}^{(+)} ({\bf p}, 0)\cr  {\bf
B}^{(+)} ({\bf p}, 0)\cr}\, , \\
\overline u_\uparrow (p^\mu)
\sigma^{\mu\nu} u_\downarrow (p^\mu) = {2im \sqrt{2} \over N} \pmatrix{
{\bf B}^{(+)} ({\bf p}, -1)\cr - {\bf E}^{(+)} ({\bf p}, -1)\cr}\, , \\
\overline u_\uparrow (p^\mu) \sigma^{\mu\nu}
v_\downarrow (p^\mu) = {2m\sqrt{2} \over N} \pmatrix{ {\bf E}^{(+)}
({\bf p}, -1)\cr  {\bf B}^{(+)} ({\bf p}, -1)\cr}\, , \\
\overline u_\downarrow (p^\mu) \sigma^{\mu\nu}
u_\downarrow (p^\mu) = -{2im \over N} \pmatrix{ {\bf B}^{(+)} ({\bf p},
0)\cr - {\bf E}^{(+)} ({\bf p}, 0)\cr}\, , \\
\overline u_\downarrow (p^\mu)
\sigma^{\mu\nu} v_\downarrow (p^\mu) = -{2m \over N} \pmatrix{ {\bf
E}^{(+)} ({\bf p}, 0)\cr  {\bf B}^{(+)} ({\bf p}, 0)\cr}\, , \\
\overline u_\downarrow (p^\mu) \sigma^{\mu\nu} u_\uparrow (p^\mu) = -
{2im \sqrt{2} \over N} \pmatrix{ {\bf B}^{(+)} ({\bf p}, +1)\cr - {\bf
E}^{(+)} ({\bf p}, +1)\cr}\, , \\
\overline u_\downarrow (p^\mu) \sigma^{\mu\nu} v_\uparrow (p^\mu) = -{2m
\sqrt{2} \over N} \pmatrix{ {\bf E}^{(+)} ({\bf p}, +1)\cr  {\bf B}^{(+)}
({\bf p}, +1)\cr}\, . \label{tcf}
\end{eqnarray}
\end{mathletters}
All the tensor currents of the type $\overline v_h (p^\mu) \sigma^{\mu\nu}
u_{h^\prime} (p^\mu)$ and $\overline v_h (p^\mu) \sigma^{\mu\nu}
v_{h^\prime} (p^\mu)$ will differ in the sign from the corresponding
expressions (\ref{tci}-\ref{tcf}) due to $\overline v_h (p^\mu) =
- \overline u_h (p^\mu) \gamma^5$ and the fact that $[\gamma^5 ,
\sigma^{\mu\nu} ]_- = 0$.

\section*{Appendix B}

Field functions of the $(1/2,1/2)$ representation (in the momentum space)
are deduced in ref.~\cite{norm} to be:
\begin{mathletters}
\begin{eqnarray}
u^\mu ({\bf p}, +1)= -{N\over
m\sqrt{2}}\pmatrix{p_r\cr m+ {p_1 p_r \over p_0+m}\cr im +{p_2 p_r \over
p_0+m}\cr {p_3 p_r \over p_0+m}\cr}&\quad&,\quad
u^\mu  ({\bf p}, -1)= {N\over m
\sqrt{2}}\pmatrix{p_l\cr m+ {p_1 p_l \over p_0+m}\cr -im +{p_2 p_l \over
p_0+m}\cr {p_3 p_l \over p_0+m}\cr}\quad,\quad\\
u^\mu ({\bf p}, 0) = {N\over m}
\pmatrix{p_3\cr {p_1 p_3 \over p_0+m}\cr {p_2 p_3 \over p_0+m}\cr m
+ {p_3^2 \over p_0+m}\cr}&\quad&, \quad
u^\mu ({\bf p}, 0_t) = {N \over m} \pmatrix{p_0\cr p_1 \cr p_2\cr
p_3\cr}\quad.
\end{eqnarray} \end{mathletters}
They do not diverge in the massless limit provided that $N=m$ and
describe the longitudinal (in the sense $h=0$) photons in this
limit.  If $N=1$
we have ``transverse photons" and the divergent behaviour of the gauge
parts of these field functions. The negative-energy potentials are
obtained by application of the complex conjugation operation (or the $CP$
conjugation, thus obtaining different field operators).

Corresponding strengths are
\begin{mathletters}
\begin{eqnarray}
{\bf B}^{(+)} ({\bf p}, +1) &=& -{iN\over 2\sqrt{2} m} \pmatrix{-ip_3 \cr
p_3 \cr ip_r\cr} =
{\bf B}^{(-)} ({\bf p}, -1) \quad,\quad      \\
{\bf B}^{(+)} ({\bf p}, 0) &=& {iN \over 2m}
\pmatrix{p_2 \cr -p_1 \cr 0\cr} = -
{\bf B}^{(-)} ({\bf p}, 0) \quad,\quad  \\
{\bf B}^{(+)} ({\bf p},
-1) &=& {iN \over 2\sqrt{2}m} \pmatrix{ip_3 \cr p_3 \cr
-ip_l\cr} =
{\bf B}^{(-)} ({\bf p}, +1)\quad,\quad
\end{eqnarray} \end{mathletters}
and
\begin{mathletters}
\begin{eqnarray}
{\bf E}^{(+)} ({\bf p}, +1) &=&  -{iN\over 2\sqrt{2} m} \pmatrix{p_0- {p_1
p_r \over p_0+m}\cr ip_0 -{p_2 p_r \over p_0+m}\cr -{p_3 p_r \over
p_0+m}\cr} = {\bf E}^{(-)} ({\bf p}, -1) \quad,\quad\\
{\bf E}^{(+)} ({\bf p}, 0) &=&  {iN \over 2m} \pmatrix{- {p_1 p_3
\over p_0+m}\cr -{p_2 p_3 \over p_0+m}\cr p_0-{p_3^2 \over
p_0+m}\cr} = - {\bf E}^{(-)} ({\bf p}, 0) \quad,\quad\\
{\bf E}^{(+)} ({\bf p}, -1) &=&  {iN\over 2\sqrt{2} m} \pmatrix{p_0- {p_1
p_l \over p_0+m}\cr -ip_0 -{p_2 p_l \over p_0+m}\cr -{p_3 p_l \over
p_0+m}\cr} = {\bf E}^{(-)} ({\bf p}, +1) \quad.
\end{eqnarray}
\end{mathletters}
They were obtained by the application of the formulas:
${\bf B}^{(\pm)} ({\bf p}, h ) =\pm {i\over 2m} {\bf p} \times {\bf
u}^{(\pm )} ({\bf p}, h)$ and ${\bf E}^{(\pm)} ({\bf p}, h) = \pm {i\over
2m} p_0 {\bf u}^{(\pm )} ({\bf p}, h) \mp {i\over 2m} {\bf p} u^{0\,
(\pm )} ({\bf p}, h)$. It is useful to compare these strengths with those
presented in~\cite[p.408]{DVA0} on using the different spinorial basis.
Corresponding cross products were also obtained in ref.~\cite{norm} and it
appears that they are related to the gauge parts ($\sim p^\mu$) of the
4-potentials in the momentum space. Relations of cross-products with
antisymmetric tensor will be given in a separate paper (preprint EFUAZ
FT-99-67 (physics/9907048), Feb. 1999).

\end{document}